\documentstyle[12pt]{article}
\newcommand\be{\begin{equation}}
\newcommand\ee{\end{equation}}
\newcommand\bea{\begin{eqnarray}}
\newcommand\eea{\end{eqnarray}}
\newcommand\ket[1]{|#1\rangle}
\newcommand\bra[1]{\langle #1|}

\newcommand{\fatalpha}{{\bf \alpha \kern -0.44em \alpha}}
\newcommand{\fatsigma}{{\bf \sigma \kern -0.54em \sigma}}
\newcommand{\tpchi}{{\bf \chi \kern -0.35em \chi}}
\newcommand{\llambda}{{\bf \lambda \kern -0.45em \lambda}}

\bibliography{plain}
\title{\bf Evaluation of relative entropy of entanglement and derivation of optimal
Lewenstein-Sanpera decomposition of Bell decomposable states via
convex optimization} \vspace{20mm}
\author{ M. A. Jafarizadeh$^{a,b,c}$
 \thanks{E-mail:jafarizadeh@tabrizu.ac.ir},
M.Mirzaee$^{a,b}$ \thanks{E-mail:mirzaee@tabrizu.ac.ir},
M.Rezaee$^{a,b}$ \thanks{E-mail:karamaty@tabrizu.ac.ir}
\\
\\
$^a${\small Department of Theoretical Physics and Astrophysics,
Tabriz University, Tabriz 51664, Iran.} \\ $^b${\small Institute
for Studies in Theoretical Physics and Mathematics, Tehran
19395-1795, Iran.} \\ $^c${\small Research Institute for
Fundamental Sciences, Tabriz 51664, Iran.}} \pagebreak

\vspace{20mm}
\begin{document}
\maketitle \vspace{15mm}
\newpage
\begin{abstract}
We provide an analytical expression for  optimal
Lewenstein-Sanpera decomposition of Bell decomposable states by
using semi-definite programming. Also using the Karush-Kuhn-Tucker
optimization method, the minimum relative entropy of entanglement
of Bell decomposable states has been evaluated and it is shown
that the same separable Bell decomposable state lying at the
boundary of convex set of separable Bell decomposable states,
optimizes both Lewenstein-Sanpera decomposition and relative
entropy of entanglement.
 {\bf Keywords: Minimum relative entropy of entanglement, Semi-definite
 programming, Convex optimization, Lewenstein-Sanpera decomposition,
Bell decomposable states.}

{\bf PACs Index: 03.65.Ud }
\end{abstract}
\vspace{70mm}
\newpage
\section{INTRODUCTION}
Quantum entanglement is a powerful property that has attracted
much attention, since it provides new means of communication,
such as quantum cryptography\cite{Ekert}, quantum
teleportation\cite{Contemp} superdense coding, and quantum
computation. Therefore, the measure of entanglement of composite
systems becomes crucial
\cite{Popescu,Scrip,vedral2,Vedral3,Bennet}. Even though, for
pure bipartite states the Von-Neumann entropy is a good measure
and it is easily calculated \cite{Popescu}; however, to quantify
the entanglement of mixed states is a harder task. Some measures
have been proposed, among them one based on relative
entropy\cite{vedral}. This measure needs a minimization
procedure\cite{Vedral3,vedral,Ramos,rel1}.\\  In order to obtain
the relative entropy of entanglement of a density matrix , the
crucial point is to find a separable state $\rho_{s}$ such that
minimizes the relative entropy of entanglement. In general case,
to find $\rho_{s}$ is a laborious work.\\ Additionally, there is
another useful tool to study entanglement, i.e.,
Lewenstein-Sanpera decomposition (LSD) \cite{LS}. Among them is
the unique optimal decomposition, the one with the largest
$\lambda$ in LSD.

From the above considerations, we see that there are some specific
separable states playing an important role in determining the
entanglement measure. Then it is meaningful to
determine the properties of these separable states.\\

On the other hand, over the past years, semidefinite programming
(SDP) has been recognized as valuable numerical tools for control
system analysis and design. In (SDP) one minimizes a linear
function subject to the constraint that an affine combination of
symmetric matrices is positive semidefinite. SDP, has been studied
(under various names) as far back as the 1940s. Subsequent
research in SDP during the 1990s was driven by applications in
combinatorial optimization\cite{Luo04}, communications and signal
processing \cite{Luo03,Luo02,Luo01}, and other areas of
engineering\cite{Luo05}. Although SDP is designed to be applied in
numerical methods it can be used for analytic computations, too.
Some authors try to use the SDP to construct an explicit
entanglement witness \cite{Doherty,Doherty1,Parrilo}. Kitaev used
SDP duality to prove the impossibility of quantum coin flipping
\cite{13}, and Rains gave bounds on distillable entanglement
using SDP \cite{17}. In the context of quantum computation,
Barnum, Saks and Szegedy reformulated quantum query complexity in
terms of SDP\cite{1}. The problem of finding the optimal
measurement to distinguish between a set of quantum states was
first formulated as a SDP in 1972 by Holevo, who gave optimality
conditions equivalent to the complementary slackness conditions
\cite{10}. Recently, Eldar, Megretski and Verghese showed that
the optimal measurements can be found efficiently by solving the
dual followed by the use of linear programming \cite{3}.
Audenaert at al. \cite{217902,032310} applied convex analysis
method and Lagrange duality  to calculate the relative entropy of
entanglement for Werner states. Also Lawrence in \cite{Lawrence}
used SDP to show that the standard algorithm implements the
optimal set of measurements. All of the above mentioned
applications indicate that the method of SDP is very useful.

 In this paper,  by using the convex optimization method, we find optimal
 LSD and  minimum relative entropy of entanglement for
BD states. Finally, we show that the separable states that
minimize relative entropy of entanglement are the same as states
which optimize the LSD which lie at the boundary of convex and
compact set of separable states \cite{Horodecki}.

The paper is organized as follows:\\ In section-2 we give a brief
review of  convex optimization. Section -3 is about the Optimal
LSD and also by using SDP we find optimal LSD for Bell
decomposable (BD) states. In section-4 we explain how one can find
the  minimum relative entropy of entanglement by using the
Karush-Kuhn-Tucker(KKT) optimization method, we find the minimum
relative entropy of entanglement of BD states. Finally it is shown
that the same separable BD state lying at the boundary of convex
set of separable BD states, optimizes both LSD and relative
entropy of entanglement. The paper is ended with a brief
conclusion.
\section{Convex Optimization}\label{semi}
\subsection{Semi-definite programming}A SDP  is a
particular type of convex optimization problem \cite{optimize}. A
SDP problem requires minimizing a linear function subject to a
linear matrix inequality (LMI) constraint \cite{Boyd}: \be
\label{sdp1}\begin{array}{cc} \mbox{minimize} & {\cal P}=c^{T}x
\\ \mbox{subject to} & F(x)\geq 0,
\end{array}\ee where $c^{T}$ is a given vector,
$x=(x_{1},...,x_{n}), $ and $F(x)=F_{0}+\sum_{i}x_{i}F_{i},$ for
some fixed hermitian matrices $F_{i}$. The inequality sign in
$F(x)\geq 0$ means that $F(x)$ is positive semidefinite.

This problem is called the primal problem. Vectors x whose
components are the variables of the problem and satisfy the
constraint $F(x) \geq 0$ are called primal feasible points, and
if they satisfy $F(x) > 0$ they are called strictly feasible
points. The minimal objective value $c^{T} x$ is by convention
denoted by ${\cal P}^{\ast}$ and is called the primal optimal
value.

Due to the convexity of  set of feasible points, SDP  has a nice
duality structure, with, the associated dual program being: \be
\label{sdp2}\begin{array}{cc} \mbox{maximize} & -Tr[F_{0}Z] \\
    & Z\geq 0 \\  & Tr[F_{i}Z]=c_{i}. \end{array}\ee

Here the variable is the real symmetric (or Hermitean) matrix Z,
and the data c, $F_{i}$ are the same as in the primal problem.
Correspondingly, matrices Z satisfying the constraints are called
dual feasible (or strictly dual feasible if $Z > 0$). The maximal
objective value of $-Tr F_{0}Z$, i.e.,  the dual optimal value, is
denoted by $d^{\ast}$.

The objective value of a primal(dual) feasible point is an upper
(lower) bound on ${\cal P}^{\ast}$($d^{\ast}$.  The main reason
why one is interested in the dual problem is that one can prove
that $d^{\ast} \leq {\cal P}^{\ast}$, and under relatively mild
assumptions, we can have ${\cal P}^{\ast} = d^{\ast}$. if the
equality holds, one can prove the following optimality condition
on x:

A primal feasible $x$ and a dual feasible $Z$ are optimal which is
denoted by $\hat{x}$ and $\hat{Z}$ if and only if  \be
\label{slacknes} F(\hat{x}) \hat{Z}=\hat{Z} F(\hat{x})=0. \ee This
latter condition is called the complementary slackness condition.

In one way or another, numerical methods for solving SDP problems
always exploit the inequality $d \leq d^{\ast} \leq {\cal
P}^{\ast} \leq {\cal P}$, where d and ${\cal P}$ are the objective
values for any dual feasible point and primal feasible point,
respectively. The difference \be\label{sdp3}{\cal
P}^{\ast}-d^{\ast}= c^{T}x+ Tr[F_{0}Z]= Tr[F(x)Z]\geq 0 \ee is
called the duality gap. If the equality holds $d^{\ast}={\cal
P}^{\ast}$ , i.e., the optimal duality gap is zero, then we say
that strong duality holds.

\subsection{Karush-Kuhn-Tucker Theorem:}
All problems treated in this paper are convex optimization
problems: minimizing a convex objective function subject to upper
bound inequality constraints on other convex functions.
  Lagrange duality theory is also well developed for convex optimization.
   For example, the duality gap is zero under mild technical conditions such as Slater ${'}$s condition
   that requires the existence of a strictly feasible
point.

 We seek first order
conditions for local optimality for the following formulation of
nonlinear programming: \be\label{kkt1}\begin{array}{cc}
\mbox{minimize} & f(x)
\\ \mbox{subject to}
    & h(x)=0 \\  & g(x)\leq 0, \end{array}
\ee

where x is a vector in $R^n$, h maps into $R^k$, and g maps into
$R^m$.

{\bf Karush-Kuhn-Tucker (KKT) Theorem:}  Assuming that functions
$g_{i}$, $h_{i}$ are differentiable and that strong duality holds,
there exists vectors $\zeta \in R^k$, and $y \in R^m$, such that
the gradient of dual Lagrangian
$L(x^{\ast},\zeta^{\ast},y^{\ast})=f(x^{\ast})+\sum_{i}\zeta^{\ast}_{i}
h_{i}(x^{\ast})+ \sum_{i} y^{\ast}_{i} g_{i}(x^{\ast})$ over x
vanishes at $x^{\ast}$ :
$$ h_{i}(x^{\ast}) = 0 \mbox{(primal feasible)}$$ $$g_{i}(x^{\ast})\leq 0
\mbox{(primal feasible)}$$ \be y_{i}^{\ast} \geq  0 \mbox{(dual
feasible)}\ee  $$y^{\ast}_{i} g_{i}(x^{\ast})=0$$
$$ \nabla f(x^{\ast})+\sum_{i}\zeta^{\ast}_{i}\nabla h_{i}(x^{\ast})+\sum_{i} y^{\ast}_{i}\nabla
g_{i}(x^{\ast})=0 $$ Then $x^{\ast}$ and
$(\zeta^{\ast},y^{\ast})$ are primal and dual optimal, with zero
duality gap. In
 summary, for any convex optimization problem with differentiable objective and
 constraint functions, any points that satisfy the KKT conditions are primal and
dual optimal, and have zero duality gap. Necessary KKT conditions
satisfied by any primal and dual optimal pair and for convex
problems, KKT conditions are also sufficient. If a convex
optimization problem with differentiable objective and constraint
functions satisfies Slater's condition, then the KKT conditions
provide necessary and sufficient conditions for optimality:
Slater's condition implies that the optimal duality gap is zero
and the dual optimum is attained, so x is optimal if and only if
there are $(\zeta^{\ast},y^{\ast})$ that, together with x,
satisfy the KKT conditions.

{\bf Slater's condition:} Suppose $x^{\ast}$ solves
\be\mbox{minimize}\;\;\;\;\; f(x)\ee
$$ g_{i}(x)\geq b_{i}\;\;,\;\; i=1,...,m
$$
 and the feasible set is non empty. Then there is a non-negative
vector $\zeta$ such that for all x

\be L(x,\zeta)=f(x)+\zeta^{T}(b-g(x))\leq
f(x^{\ast})=L(x^{\ast},\zeta). \ee If in addition $f(.),
g_{i}(.)\;\;,\;\;i=1,...,m$ are continuously differentiable
\be\frac{\partial f}{\partial x_{j}}(x^{\ast})- \zeta
\frac{\partial g}{\partial x}(x^{\ast})=0.\ee

 In the spatial case the  vector $x^{\ast}$ is a
solution of the linear program $$ \mbox{minimize}\;\;\;\; c^{T}x$$
\be\mbox{subject to} \;\;\;\;Ax = b\ee
$$\;\;\;\;\;\;\;\;\;\;\;\;\;\;\;\;\;\;\; x \geq 0,
$$

if and only if there exist vectors $\zeta \in R^k$, and $y \in
R^m$ for which the following conditions hold for
$(x,\zeta,y)=(x^{\ast},\zeta^{\ast},y^{\ast})$
 $$ A^{T}\zeta +
y=c$$ \be\label{tkk} Ax=b\ee $$x_i \geq 0;\;\; y_i \geq 0; \;\;
x_{i}y_{i}= 0\;\; ,\;\; i = 1,...,m.$$ A solution
$(x^{\ast},\zeta^{\ast},y^{\ast})$ of (\ref{tkk}) is called
strictly complementary, if $x^{\ast}+y^{\ast} > 0$, i. e., if
there exists no index $i \in \{1,...,m\}$ such that
$x_{i}^{\ast}=y_{i}^{\ast}=0$.
\section{Lewenstein-Sanpera decomposition}\setcounter{equation}{0}
According to LSD \cite{LS}, any bipartite  density matrix $\rho$
can be written as
\begin{equation}\label{LSD}
\rho=\lambda\rho_{s}+(1-\lambda)\rho_e, \quad\quad
\lambda\in[0,1],
\end{equation}
where $\rho_{s}$ is a separable density matrix and $\rho_e$ is an
entangled state. The LSD of a given density matrix $\rho$ is not
unique and, in general, there is a continuum set of LSD to choose
from. The decomposition with the largest weight $\lambda$ of the
separable part is the optimal LSD \cite{akhtar,englert}, which
can be determined uniquely.

Furthermore, in the case of a two-qubit state it can be shown
\cite{LS} that the entangled part $\rho_e$ of the optimal LSD is
always a pure state i.e, \be
 \rho=(1-\lambda)\ket{\psi}\bra{\psi}+\lambda\rho^{'}_{s}\ee

\subsection{Optimal Lewenstein-Sanpera decomposition for
Bell-decomposable states via convex optimization} A BD state is
defined by
\begin{equation}
\rho=\sum_{i=1}^{4}p_{i}\left|\psi_i\right>\left<\psi_i\right|,\quad\quad
0\leq p_i\leq 1,\quad \sum_{i=1}^{4}p_i=1,
 \label{BDS1}
\end{equation}
where $\left|\psi_i\right>$ are Bell states given by
\begin{eqnarray}\nonumber
\label{BS12} \left|\psi_1\right>=\left|\phi^{+}\right>
=\frac{1}{\sqrt{2}}(\left|00\right>+\left|11\right>), \qquad
\nonumber \left|\psi_2\right>=\left|\phi^{-}\right>
=\frac{1}{\sqrt{2}}(\left|00\right>-\left|11\right>), \\
\label{BS34}\left|\psi_3\right>=\left|\psi^{+}\right>
=\frac{1}{\sqrt{2}}(\left|01\right>+\left|10\right>), \qquad
\left|\psi_4\right>=\left|\psi^{-}\right>
=\frac{1}{\sqrt{2}}(\left|01\right>-\left|10\right>).
\end{eqnarray}
In terms of Pauli's matrices, $\rho$ can be written as,

\begin{equation}
\rho=\frac{1}{4}(I\otimes I+\sum_{i=1}^{3}
t_i\sigma_{i}\otimes\sigma_{i}), \label{BDS2}
\end{equation}
where
\begin{equation}\label{t-p}
\begin{array}{l}
t_1=p_1-p_2+p_3-p_4, \\
t_2=-p_1+p_2+p_3-p_4, \\
t_3=p_1+p_2-p_3-p_4.
\end{array}
\end{equation}
From positivity of $\rho$ we get
\begin{equation}\label{T1}
\begin{array}{ll}
1+t_1-t_2+t_3\geq 0, & \qquad 1-t_1+t_2+t_3\geq 0, \\
1+t_1+t_2-t_3\geq 0, & \qquad 1-t_1-t_2-t_3\geq 0.
\end{array}
\end{equation}
These equations form a tetrahedral  with its vertices located at
$(1,-1,1)$, $(-1,1,1)$, $(1,1,-1)$, $(-1,-1,-1)$ \cite{horo2}. In
fact these vertices denote the Bell states  given in (\ref{BS34}).

On the other hand  $\rho$ given in (\ref{BDS2}) is separable if
and only if $t_i$ satisfy Eq. (\ref{T1}) and,
\begin{equation}\label{T2}
\begin{array}{ll}
1+t_1+t_2+t_3\geq 0, & \qquad 1-t_1-t_2+t_3\geq 0, \\
1+t_1-t_2-t_3\geq 0, & \qquad 1-t_1+t_2-t_3\geq 0.
\end{array}
\end{equation}

inequalities (\ref{T1}) and (\ref{T2}) form an octahedral with its
 vertices located at
$O_1^{\pm}=(\pm 1,0,0)$, $O_2^{\pm}=(0,\pm 1,0)$ and
$O_3^{\pm}=(0,0,\pm 1)$. So, tetrahedral is divided into five
regions. Central regions, defined by octahedral, are separable
states ($p_{k}\leq \frac{1}{2}$). There are also four smaller
equivalent tetrahedral corresponding to entangled states($p_{k}>
\frac{1}{2}$ for only one of $k=1,...,4$), where
$p_{k}=\frac{1}{2}$ denote to boundary between separable and
entangled region. Each tetrahedral takes one Bell state as one of
its vertices (see Fig-1).

Now in order to obtain optimal LSD via convex optimization of
entangled BD state given in (\ref{BDS1}), with
$p_{1}>\frac{1}{2}$, we first choose an arbitrary separable state
as
 \be\label{ICDS3}
\rho^{'}_{s}=\sum_{i=1}^{4}p^{'}_{i}\left|\phi_i\right>\left<\phi_i\right|,\quad\quad
\sum_{i=1}^{4}p^{'}_i=1,\;\;\;p^{'}_1<\frac{1}{2}\ee in the
separable region. Then according to strict  SDP optimization
prescription of section (2), we need  to optimize
$Tr(\Lambda\rho_{s}^{\prime})$ with respect to
$\rho-\Lambda\rho_{s}^{\prime} > 0$, where the feasible solution
corresponds to \be \label{ICDS4}
\Lambda_{max}=min\{\frac{p_{1}}{p_{1}^{\prime}},\frac{p_{2}}{p_{2}^{'}},\frac{p_{3}}{p_{3}^{'}},\frac{p_{4}}{p_{4}^{'}}
\}. \ee One can show that the only possible choice of
$\Lambda_{max}$ consistent with positivity of $\rho
-\Lambda_{max}\rho_{s}^{\prime}$ is \be \label{BD8} \Lambda_{max}=
\frac{(1-p_{1})}{(1-p^{'}_{1})}\;\;\;\;\; \mbox{and}
\;\;\;p^{\prime}_{i}=\frac{p_{i}}{\Lambda_{max}}\;,i=2,3,4.\ee

The equation (\ref{BD8}) indicates that $\Lambda_{max}$ is a
monotonic increasing function of $p^{\prime}_{1}$ and its maximum
value corresponds to $p_{1}^{\prime}=\frac{1}{2}$, with \be
\Lambda_{max}=2(1-p_{1})\;\;\;\; \mbox{and} \;\;\;
p^{\prime}_{i}=\frac{p_{i}}{2(1-p_{1})}\;,i=2,3,4,\ee  where this
separable state lies at the  boundary of separable region, see
Fig-1. Substituting the results that obtained  in
$\rho-\Lambda_{max}\rho_{s}^{\prime}$ we get
\be\rho-\Lambda_{max}\rho^{'}_{s}=(2p_{1}-1)\ket{\phi_{1}}\bra{\phi_{1}},\ee
which is a pure states in agreement with theorem (2) of
Ref.\cite{LS}. Equation (3-13) with $\Lambda_{max}$ given in
(3-12) is nothing but  the optimal LSD of BD states.

\section{Evaluation of relative entropy of entanglement}
\subsection{Minimizing relative entropy via KKT method}
Fundamental to our understanding of distinguishability is the
measure of uncertainty in a given probability distribution. The
uncertainty in a collection of a probability mass function (pmf)
$\omega\in {\cal R}^k$ is given by its entropy,
\begin{equation}
S(p)=-\sum_{i}\omega_{i}\log (\omega_{i})
\end{equation}
called the Shannon entropy.  This measure is suitable for the
states of systems described by the laws of classical physics, but
it will have to be changed, along with other classical measures,
when we present the quantum information theory. We ultimately
wish to be able to talk about storing and processing information.
For this we require a means of comparing two different
probability distributions, which is why we introduce the notion
of relative entropy (first introduced by Kullback and
Leibler, 1951 \cite{Leibler}).\\
 {\bf Definition of the relative entropy:}

Suppose that we consider  the problem of estimating a probability
mass function (pmf) $\omega\in {\cal  R}^k$ given a strictly
positive prior $q\in {\cal  R}^k$. The relative entropy between
these two distributions is defined as \cite{vedral}
\begin{equation}\label{ent1}
 {\cal I}(\omega;q)=\sum_{i=1}^{k}\omega_{i}\log\frac{\omega_{i}}{q_{i}}
\end{equation}
This function is a measure of the 'distance' between $\omega_{i}$
and $q_{i}$, even though, strictly speaking, it is not a
mathematical metric since it fails to be symmetric $ {\cal
I}(\omega_{i}\mid\mid q_{j})\neq {\cal I}(q_{j}\mid\mid
\omega_{i})$.

A common restriction on $\omega$ is a mean constraint, so that if
there are k observations $A = [a_{1}, . . . , a_{k}] \in  {\cal
R}^{d×k}$ and a mean $b \in  {\cal R}^{d}$, then $\omega$ must
satisfy $A\omega = b$. A standard approach
\cite{Shore,Kullback,Csisz} is to minimize the relative entropy
function (\ref{ent1}) over the constrained probability simplex
\begin{equation}\label{slater}
\begin{array}{cc}  1^{T}\omega=1, & \omega\geq 0 \\
A\omega=b, & \end{array}\end{equation} where the first two
constraints as due to the fact that $\omega$ represents a
distribution and the symbol $ 1$ denotes a vector of ones.
Consider the relative entropy optimization problem
\begin{equation}\label{slater1} \mbox{minimize}\;\;\;{\cal I}(\omega;q)\end{equation} subject to constraint (\ref{slater}).
The optimization problem (\ref{slater1}) is convex; in other words
its objective function is convex, and the equality constraints
are linear. Under a suitable constraint qualification, such as
Slater's condition, the first-order optimality conditions of
(\ref{slater1}) are in fact both necessary and sufficient (see,
for example, \cite{Boyd,Mangasarian}).

One simple constraint qualification is Slater's condition, i.e.,
there exists an $\omega$ in the relative interior of the feasible
set, such that
\begin{equation}\label{opt1}
\begin{array}{ccc} \omega> 0,  & 1^{T}\omega=1, & A\omega=b.
\end{array}\end{equation}
Let $\zeta$ and $y$ be the Lagrange multipliers associated with
the first and second constraints of (\ref{slater1}), respectively.

Definition: (KKT optimality conditions): A triple
($\omega^{\ast}$, $\zeta^{\ast}$, $y^{\ast}$) is a first-order KKT
point of (\ref{slater1}) if it satisfies the following conditions:
\begin{equation}\label{opt2}\begin{array}{cc} 1^{T}\omega=1 \\ A\omega=b \\
\nabla_{\omega}{\cal I}(\omega,q)+\zeta
1+A^{T}y=0.\end{array}\end{equation}Theorem:  Suppose that
Slater's condition holds. Then the vector with components
\begin{equation}\label{opt3}
\omega^{\ast}_{j}=\frac{u_{j}}{\sum_{j=1}^{k}u_{j}},
\end{equation}
for $j = 1, . . . , k$, solves (\ref{slater1}), where
\begin{equation}\label{opt4}
u_{j}=q_{j}\exp(-a_{j}^{T}y^{\ast}),
\end{equation}
and $y^{\ast}$ is the Lagrange multiplier corresponding to the
constraint $A\omega = b$ \cite{rel1}. \subsection{Evaluation of
relative entropy of entanglement of Bell decomposable states via
KKT method} In order to obtain minimal relative entropy of
entanglement we choose entangled BD state (given in (\ref{BDS1})),
with $p_{1}>\frac{1}{2}$ and then we choose an arbitrary separable
states \be\label{ICDS3}
\rho^{'}_{s}=\sum_{i=1}^{4}p^{'}_{i}\left|\phi_i\right>\left<\phi_i\right|,\quad\quad
0\leq p^{\prime}_i\leq
1,\;\;\sum_{i=1}^{4}p^{'}_i=1,\;\;\;p^{'}_1=\frac{1}{2},\ee at
the boundary of  separable region.

Then according to the prescription  of the minimization of the
relative entropy given in  section (2), all we need is to
minimize ${\cal I}(\omega;q)$ with
$\omega^{T}=({p^{\prime}}_{1},{p^{\prime}}_{2},{p^{\prime}}_{3},{p^{\prime}}_{4})$,
$q^{T}=(p_{1},p_{2},p_{3},p_{4})$ , $p^{\prime}_{1}\leq
\frac{1}{2}$ and   $p_{1}>\frac{1}{2}$ subject to the constraints
(\ref{slater})with matrix A and vector b of the following form:
\be\label{op1}\begin{array}{ccc}A=\left(\begin{array}{cccc} 1 & 0
& 0 & 0
\\ 0 & 0 & 0 & 0 \\ 0 & 0 & 0 & 0 \\ 0 & 0 & 0 & 0
\end{array}\right) & , & b^{T}=\left(\begin{array}{cccc}b_{1} & 0 &0 &0
\end{array}\right)\end{array} ,\ee
where $b_{1}<\frac{1}{2}$. Now, using the solutions (\ref{opt3})
and (\ref{opt4}) of KKT (optimality conditions) or relative
entropy optimization given in (\ref{opt2}) we obtain
\be\label{op3}\begin{array}{ccccccc}u_{1}=p_{1}\exp(-y_{1}^{\ast})&
, & u_{2}=p_{2} & , & u_{3}=p_{3} & , & u_{4}=p_{4}\end{array},
\ee now, using the
 slater's constraint  we obtain \be\label{opt7}
\omega^{\ast}_{1}=\frac{p_{1}\exp(-y_{1}^{\ast})}{p_{1}\exp(-y_{1}^{\ast})+p_{2}+p_{3}+p_{4}}=b_{1}.\ee
After some calculation one can find the Lagrange multiplier
$y_{1}^{\ast}$ and the other components of optimal
$\omega^{\ast}$, where we quote only the optimal values
$\omega^{\ast}_{j}\;\;,\;\; j=2,3,4$ in the following:

\be\label{opt8}\begin{array}{ccccccc}\omega^{\ast}_{1}=b_{1}&,&\omega^{\ast}_{2}=(\frac{1-b_{1}}{1-p_{1}})p_{2}&,&\omega^{\ast}_{3}=(\frac{1-b_{1}}{1-p_{1}})p_{3}&,&\omega^{\ast}_{4}=(\frac{1-b_{1}}{1-p_{1}})p_{4}\end{array}.
\ee Finally inserting to optimal $\omega^{\ast}$ in (\ref{opt8}),
we obtain the   following expression for the relative entropy:
\be\label{opt9} {\cal
I}(\omega^{\ast};q)=\omega^{\ast}_{1}\log\frac{\omega^{\ast}_{1}}{p_{1}}+\omega^{\ast}_{2}\log\frac{\omega^{\ast}_{2}}{p_{2}}+\omega^{\ast}_{3}\log\frac{\omega^{\ast}_{3}}{p_{3}}+\omega^{\ast}_{4}\log\frac{\omega^{\ast}_{4}}{p_{4}}
=b_{1}\log(\frac{b_{1}}{p_{1}})+(1-b_{1})\log(\frac{1-b_{1}}{1-p_{1}}).\ee
The equation (\ref{opt9}) indicates that ${\cal I}(\omega;q)$ is a
monotonic increasing function of $b_{1}$ and its maximum value
corresponds to $p_{1}^{\prime}=b_{1}=\frac{1}{2}$, with \be {\cal
I}(\omega;q)=\frac{1}{2}\log(\frac{1}{2p_{1}})+\frac{1}{2}\log(\frac{1}{2(1-p_{1})}).\ee
and \be\label{opt12}
p^{\prime}_{i}=\omega^{\ast}_{i}=\frac{p_{i}}{2(1-p_{1})}\;,i=2,3,4.\ee
The BD separable state with
$p^{\prime}_{i}=\omega^{\ast}_{i}=\frac{p_{i}}{2(1-p_{1})}$ which
minimizes the relative entropy of entanglement of entangled BD
state $\rho=\sum_{i=1}^{4}p_{i}<{\psi}\mid{\psi}>$ with
$p_{1}>\frac{1}{2}$ is the same as the one that has been found by
Vedral and Plenio in Ref. \cite{vedral2}. Also one should
notice   that
 $c=2p_{1}-1$, where c is
concurrence of BD states and we have \be\label{opt10} {\cal
I}(\omega;q)= -\frac{1}{2}\log{(1-c^{2})},\ee which is in
agreement with Ref.\cite{Vedral3}.\\Finally comparing the results
of sections 3 and 4 we see that  the same separable BD state lying
at the boundary of convex set of separable BD states, optimizes
both LSD  and relative entropy of entanglement.
\section{conclusion}
Here in this work, using the convex optimization method, we have
been able to obtain optimal LSD of BD states ( by using the
semidefinite programming method) and we have also evaluated the
relative entropy of entanglement of BD states with respect to
convex set of BD separable states ( by using KKT method).

At the end it is shown that the same separable BD state lying at
the boundary of convex set of separable BD states, optimizes both
LSD and relative entropy of entanglement.

\newpage

\vspace{10mm}

{\Large {\bf Figure Captions}}

\vspace{10mm}

Figure-1: All BD states are defined as points interior to
tetrahedral. Vertices $P_{1}$, $P_{2}$, $P_{3}$ and $P_{4}$ denote
projectors corresponding to Bell states given in Eqs. (\ref{BS34})
. Interior points of Octahedral correspond to separable states.
\vspace{10mm}
\end{document}